\UseRawInputEncoding

\documentclass[journal,final]{IEEEtran}
\usepackage{amssymb}
\usepackage{amsmath}
\usepackage{xcolor}
\ifCLASSINFOpdf
\else
   \usepackage[dvips]{graphicx}
\fi
\usepackage{url}

\usepackage{pifont}
\newcommand{\cmark}{\ding{51}}%
\newcommand{\xmark}{\ding{55}}%

\usepackage[numbers,sort&compress]{natbib}

\usepackage{graphicx}
\usepackage{multirow}

\usepackage[]{review}
\setcoverletter{coverletter-R1.tex}
\setrevision{1}

\begin{document}

\title{Multichannel Long-Term Streaming Neural Speech Enhancement for Static and Moving Speakers}

\author{Changsheng Quan, Xiaofei Li
\thanks{Changsheng Quan is with Zhejiang University, Hangzhou 310058, China, and also with the School of Engineering, Westlake University, Hangzhou 310030, China. E-mail: quanchangsheng@westlake.edu.cn}
\thanks{Xiaofei Li is with the School of Engineering, Westlake University, Hangzhou 310030, China, and also with the Institute of Advanced Technology, Westlake Institute for Advanced Study, Hangzhou 310024, China. Corresponding author: lixiaofei@westlake.edu.cn.}}

\markboth{Journal of \LaTeX\ Class Files, Vol. 14, No. 8, August 2015}
{Shell \MakeLowercase{\textit{et al.}}: Bare Demo of IEEEtran.cls for IEEE Journals}
\maketitle

\begin{abstract}

In this work, we extend our previously proposed offline SpatialNet for long-term streaming multichannel speech enhancement in both static and moving speaker scenarios. SpatialNet exploits spatial information, such as the spatial/steering direction of speech, for discriminating between target speech and interferences, and achieved outstanding performance. The core of SpatialNet is a narrow-band self-attention module used for learning the temporal dynamic of spatial vectors. Towards long-term streaming speech enhancement, we propose to replace the offline self-attention network with online networks that have linear inference complexity w.r.t signal length and meanwhile maintain the capability of learning long-term information. Three variants are developed based on (i) masked self-attention, (ii) Retention, a self-attention variant with linear inference complexity, and (iii) Mamba, a structured-state-space-based RNN-like network. 
Moreover, we investigate the length extrapolation ability of different networks, namely test on signals that are much longer than training signals, and propose a short-signal training plus long-signal fine-tuning strategy, which largely improves the length extrapolation ability of the networks within limited training time.  
Overall, the proposed online SpatialNet achieves outstanding speech enhancement performance for long audio streams, and for both static and moving speakers. The proposed method is open-sourced in https://github.com/Audio-WestlakeU/NBSS.

\end{abstract}


\IEEEpeerreviewmaketitle

\section{Introduction}

\IEEEPARstart{M}{ultichannel} speech enhancement techniques have been investigated broadly in the era of deep learning \cite{hershey_DeepClusteringDiscriminative_2016,yu_permutation_2017,chen_BeamGuidedTasNetIterative_2022,wang_TFGridNetIntegratingFull_2022}. 
Most of the methods in the literature \addnote[assume]{1}{assume} that the speakers are static and process relatively short signal segment.
This work aims to design a neural network for multichannel long-term streaming speech denoising and dereverberation in both moving and static speaker scenarios.

Multichannel speech enhancement algorithms can be broadly categorized into three classes: conventional methods, two-stage methods and end-to-end methods. Conventional methods like weighted prediction error (WPE) \cite{nakatani2010speech} and beamforming \cite{gannot_consolidated_2017} mainly leverage spatial information, including signal propagation and sound field, to enhance signals.
Two-stage methods extend conventional methods, especially beamforming, by using DNNs to aid the estimation of beamforming parameters, such as the spatial covariance matrices (SCM) of desired and undesired signals \cite{heymann_blstm_2015, ochiai_beam-tasnet_2020,  chen_BeamGuidedTasNetIterative_2022,  gu_UnifiedAllNeuralBeamforming_2023, wang_attention-driven_2023, ochiai_MaskBasedNeuralBeamforming_2023}. End-to-end methods \cite{tesch_InsightsDeepNonLinear_2023,quan_spatialnet_2024,luo_DualPathRNNEfficient_2020,luo_conv-tasnet_2019,yang2022mcnet} directly estimate target signals using DNNs. Online speech enhancement networks are proposed in \cite{wang_attention-driven_2023, ochiai_MaskBasedNeuralBeamforming_2023, yang2022mcnet}.

In \cite{quan_spatialnet_2024}, we have proposed an offline multichannel speech enhancement and separation network, called SpatialNet, which achieved the state-of-the-art performance on all the six experimental datasets. Just as its name implies, SpatialNet is designed to learn spatial information for discriminating between target speech and interferences, such as the difference of spatial correlation between directional speech and diffuse ambient noise, the convolutional property of reverberation, and the different steering directions/vectors of different speakers. 
SpatialNet performs end-to-end speech enhancement in the short-time Fourier transform (STFT) domain, with interleaved narrow-band blocks and cross-band blocks, which are responsible for learning the narrow-band temporal-spatial information and the cross-band correlation of narrow-band spatial representation, respectively. The narrow-band block processes the data sequence of each frequency independently, using a regular self-attention (SA) module and a convolutional-augmented feed-forward module. The cross-band blocks process time frames independently. 

In this work, we extend the offline SpatialNet to be an online network that is not only able to inherit the strong capability of SpatialNet but also computationally efficient for processing very long audio streams. The extension of the convolutional layers (modified as causal convolutions) and the cross-band blocks (kept unchanged) are straightforward. However, it is nontrivial to extend the narrow-band SA module, which learns long-term temporal-spatial information, and plays the most important role for the success of SpatialNet as presented in \cite{quan_spatialnet_2024}. For online/streaming processing, the offline SA module should be modified to i) be causal, ii) has a linear complexity w.r.t signal length, in contrast to the regular SA that has a quadratic complexity, 
and iii) has comparable capability as regular SA in terms of learning long-term information. To satisfy these conditions, we propose three variants of online SpatialNet: i) masked SA \cite{vaswani_attention_2017} simply restricts the self-attention to a constant number of past time steps, ii) Retention \cite{sun_retentive_2023} is a recently proposed linearized self-attention variant, iii) Mamba \cite{gu_mamba_2023} is a recently proposed structured state space sequence model (SSM). In the framework of SpatialNet, the three variants are carefully developed for online speech enhancement in both static and moving speaker scenarios. 
Moreover, in this work, we also investigate the length extrapolation ability of different networks, which is rarely discussed in the previous online speech enhancement works \cite{wang_attention-driven_2023, ochiai_MaskBasedNeuralBeamforming_2023, yang2022mcnet}, 
and we propose a short-signal training plus long-signal fine-tuning (ST+LF) strategy, which largely improves the length extrapolation ability of the networks using limited training time.  
Overall, combining the advanced SpatialNet architecture, the advanced streaming networks (Retention and Mamba), and the ST+LF strategy, the proposed online SpatialNets achieve outstanding online speech enhancement performance for long audio streams, and for both static and moving speakers.

\vspace{-0.3cm}
\section{Online SpatialNet}
This section presents the proposed online SpatialNet (oSpatialNet) as an extension of our previously proposed offline SpatialNet \cite{quan_spatialnet_2024}. 
SpatialNet first uses an input convolution layer to conducts convolution \addnote[tfconv]{1}{with a kernel size of 1 and 5 on the frequency and time axes respectively} of the multichannel microphone signals ${\mathbf{x}}[f,t,:]\in \mathbb{R}^{2M}$, where $f$, $t$ and $2M$ are the frequency index, time frame index, and the number of (real and imaginary) channels, and outputs corresponding hidden units $\textbf{h}[f,t,:]\in \mathbb{R}^H$, where $H$ is hidden dimension for representing each T-F bin. Then $L$ interleaved narrow-band and cross-band blocks process the hidden units. Finally, the STFT coefficients of target speech (the direct-path speech) are estimated from $\textbf{h}[f,t,:]$ by an output linear layer. The narrow-band block processes frequencies independently, and regular SA and time-convolutional feed forward network (T-ConvFFN) are used for modeling the rich temporal-spatial information in one frequency \cite{li_MultichannelOnlineDereverberation_2019,yoshioka_GeneralizationMultiChannelLinear_2012, nakatani_UnifiedConvolutionalBeamformer_2019, gannot_consolidated_2017, winter_map-based_2006, boeddecker_front-end_2018} in an offline way. The cross-band block processes time frames independently, which will be kept unchanged and not described in this work,  
please refer to \cite{quan_spatialnet_2024} for more details.
The network of interleaved cross-band block and narrow-band block is shown in Fig. \ref{fig_spatialnet_online} (a).


\vspace{-0.3cm}
\subsection{Online SpatialNet}

\begin{figure}[tbp]
  \centering
  \includegraphics[width=0.9\linewidth]{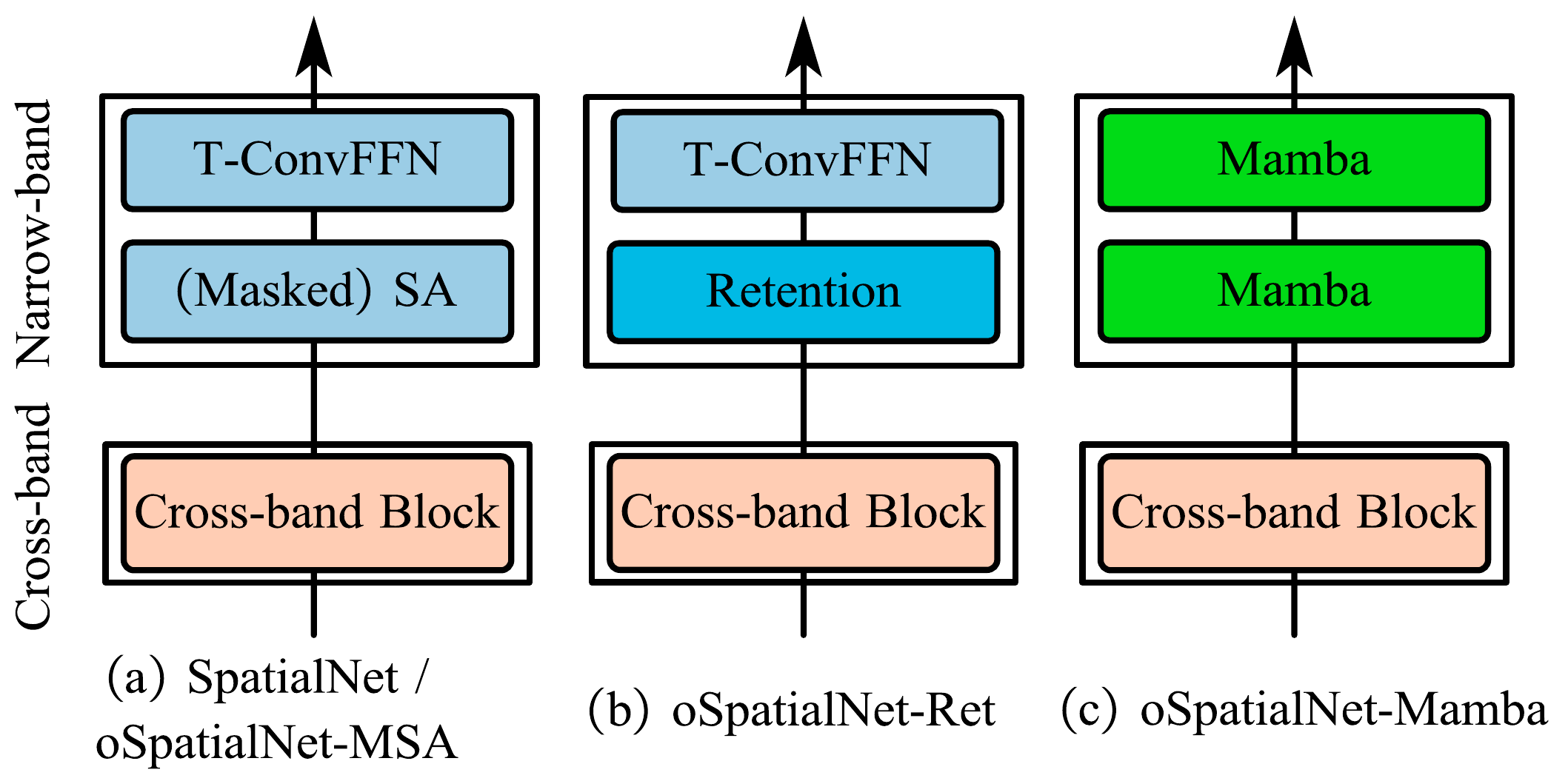}
  \vspace{-0.3cm}
  \caption{Network structure of SpatialNet and three variants of oSpatialNet.}
  \label{fig_spatialnet_online}
  \vspace{-0.5cm}
\end{figure}

In oSpatialNet, the temporal-convolutional layers in the narrow-band block are modified to be causal if they are used. Three streaming network architectures, namely masked SA \cite{vaswani_attention_2017}, Retention \cite{sun_retentive_2023} and Mamba \cite{gu_mamba_2023}, are used to redesign the narrow-band block.
The three variants are referred to as oSpatialNet-MSA, oSpatialNet-Ret and oSpatialNet-Mamba, respectively, and their structures are presented in Fig. \ref{fig_spatialnet_online}.

\subsubsection{Masked SA} As a direct and widely used approach, masked SA (MSA) uses the time-restrict masking \cite{vaswani_attention_2017, zhang_transducer_2020, moritz_streaming_2020} for streaming processing:
\begin{equation}
    \textbf{h}[f,t,:] \leftarrow \text{MSA}(\textbf{h}[f,t,:],\textbf{h}[f,t-l:t-r,:],\textbf{h}[f,t-l:t-r,:]) \notag
\end{equation}
where the first, second and third parameters of MSA are respectively the query, key and value vectors. $l$ and $r$ ($r=0$ in this paper) are the number of restricted past and future frames that are used for steaming self-attention, respectively. 


\subsubsection{Retention}
As a linearized attention variant, i.e. \addnote[italicO]{1}{\textit{O}(1)} inference cost for each time step, retention compresses all historic context into one state matrix $S_{t-1} \in \mathbb{R}^{H\times H}$, then for a new time step $t$, the state matrix $S_t$ is generated using the key vector $k_t \in \mathbb{R}^{H\times 1}$ and value vector $v_t \in \mathbb{R}^{H\times 1}$ as:
\begin{equation}
    S_t = \gamma S_{t-1} + k_t v_t^{\mathsf{T}} 
    \label{eq_retention_state}
\end{equation}
where $\gamma \in [0,1]$ is the decay weight for the previous state, and $^{\mathsf{T}}$ denotes transpose. The output is queried from the state:
    $o_t = q_t^{\mathsf{T}} S_t$, 
where $q_t\in \mathbb{R}^{H\times 1}$ is the query vector for time step $t$.
Same with self-attention, $q_t$, $k_t$ and $v_t$ are transformed from the layer input, i.e. $\mathbf{h}[f,t,:]$, by linear projections, and $o_t$ is processed to update $\mathbf{h}[f,t,:]$ by a linear projection. In retention, the rotary positional encoding (RoPE) \cite{su_roformer_2024,sun_length-extrapolatable_2023} is applied on $q_t$ and $k_t$ after the linear projections. We found that the RoPE is not helpful for our speech enhancement task, and may cause the length extrapolation problem, so we decided to remove it. 

\subsubsection{Mamba}
Structured state space sequence models (SSMs) are inspired by the classic continuous-time state space models, in which the relationship between its input $x(\tau)\in \mathbb{R}$, state $s(\tau) \in \mathbb{R}^{N}$ and output $y(\tau) \in \mathbb{R}$ can be formulated as \cite{gu_mamba_2023}:
\begin{align}\label{eq_continuous_obs}
    s{'}(\tau) = {A}s(\tau) + {B}x(\tau), \quad y(\tau) = {C}s(\tau)  
\end{align}
where $\tau$ is the continuous-time index, $A \in\mathbb{R}^{N\times N}$, $B\in\mathbb{R}^{N}$, $C\in\mathbb{R}^{N}$ are the system parameters, and $s{'}(\tau)$ is the derivative of $s(\tau)$ with respect to $\tau$. This continuous-time differential system can be discretized into a discrete-time system,  
\begin{equation}
\begin{aligned}\label{eq_discrete_state}
    s(t) = \overline{A}s(t-1) + \overline{B}x(t), \quad y(t) = Cs(t)      
\end{aligned}
\end{equation}
where $\overline{A}\in\mathbb{R}^{N\times N}$ and $\overline{B}\in\mathbb{R}^{N}$ are derived from  $A$ and $B$ based on certain discretization rules. In SSMs, such discrete-time state model is independently applied to each hidden unit. In mamba \cite{gu_mamba_2023}, the input-independent $\overline{A}$, $\overline{B}$ and $C$ are replaced by input-dependent $\overline{A}(t)$, $\overline{B}(t)$ and $C(t)$ for efficiently selecting informative data in a sequence.
The main structure of mamba is composed of three linear layers, one depth-wise CNN (DW-CNN) and one selective SSM:
\begin{equation}
\begin{aligned}
    \textbf{h}{'}[f,:,:] \leftarrow & \ \text{SSM(SiLU(DW-CNN(Linear(}\textbf{h}[f,:,:])))) \\ 
    & \odot \text{SiLU(Linear(}\textbf{h}[f,:,:])) \in \mathbb{R}^{T \times 2H} \\ \notag
    \textbf{h}[f,:,:] \leftarrow & \ \text{Linear(}\textbf{h}{'}[f,:,:]) \in \mathbb{R}^{T\times H}  \notag
\end{aligned}
\end{equation}
where $\odot$ denotes element-wise multiplication, the DW-CNN and selective SSM both process $\textbf{h}$ along the time axis. 
As shown in Fig. \ref{fig_spatialnet_online} (c), the SA and T-ConvFFN together is replaced with two mamba blocks following \cite{gu_mamba_2023}.


\vspace{-0.3cm}
\subsection{Training Strategy: Short Training plus Long Fine-tuning}
\addnote[most]{1}{In most} streaming speech enhancement applications, the model is required to process very long audio streams. However, the majority of current \addnote[research]{1}{academic research} in the field only focus on processing short segments of a few seconds to a dozen seconds. In this work, we investigate how to handle long signals. 
The common practice  is to train the networks with short signals and directly test on long signals, as training with long signals will (such as quadratically for self-attention) increase the training complexity. However, short-signal training could be problematic in terms of (i) learning very-long-term temporal dependencies. The proposed oSpatialNet mainly exploits  spatial information for discriminating between target speech and interferences. 
Spatial information could be invariant over very long time, for example the spatial correlation of directional speech and diffuse ambient noise, and the RIRs for static speakers; (ii) causing the length extrapolation problem, namely the performance could collapse when the test sequence is much longer than the training sequences. 

Training with long signals is \addnote[beneficial]{1}{beneficial} for resolving these problems. The networks used in the proposed model, i.e. Retention and Mamba, have a linear complexity w.r.t signal length, which provides a good foundation for long-signal training. Even though, the computational time is still very high when training with very-long signals. 
To solve this problem, we propose a training strategy called ``short-signal training plus long-signal fine-tuning" (ST+LF). Specifically, in our experiments, we first train the networks using short speech segments (e.g., 4 s) for a large number of epochs to fully learn the short-term speech enhancement knowledge, and then 
fine-tune the network using longer speech segments (e.g., 32 s) for a few epochs to extend the knowledge to long-term processing. 
This strategy is efficient in terms of both training cost and speech enhancement performance. 

\section{Experiments}
\vspace{-0.0cm}
\subsection{Dataset}
The proposed method is evaluated on simulated datasets with both static and moving speakers. 
Rooms are simulated with the length, width and height sampled in [4, 10] m, [4, 10] m and [3, 4] m, respectively. The reverberation time (RT60) for each room is sampled from [0.1, 1.0] s. A 6-channel microphone array that has the same topology as the CHiME-3 microphone array \cite{barker_ThirdCHiMESpeech_2015} is simulated, and is located in a square area with a length of 1 m in the center of the room. 
40000, 4000 and 4000 loop trajectories are simulated \addnote[gpurir]{1}{using gpuRIR \cite{diaz-guerra_gpurir_2021}} in randomly sampled rooms for training, validation and test, respectively. 
The clean speech signals from the WSJ0 corpus are used to generate the reverberant speech. For moving speakers, the start location and moving direction are randomly sampled in the loop trajectories. The moving speed is sampled in [0.12, 0.4] m/s \addnote[speed]{1}{following \cite{taseska_BlindSourceSeparation_2018}}. 
Note that, when moving speed is high, the simulated direct-path target signals may have some clicking noise and other audible artifacts \cite{lehmann_diffuse_2010}. 
To mitigate this problem, the speech signals at two adjacent locations are partial-overlapped and applied with a trapezium window for both the reverberant signal and direct-path signals.
For static speakers, speaker locations are randomly sampled from the loop trajectories.
The real-recorded noise signals of the CHiME-3 dataset \cite{barker_ThirdCHiMESpeech_2015} are added with a signal-to-noise ratios sampled from [-5, 10] dB. 
The moving- and static-speaker signals are jointly used for training networks and they have equal proportions in the training set.
\addnote[train-test-diff]{1}{The training set and test set are generated in the same way, but they have different signal length.}
\addnote[sf]{1}{The sampling rate is set to 8 kHz, to have a lower training cost and higher development speed. Our preliminary experiments demonstrated that the models will behave similarly when a higher sampling rate is used.}


\vspace{-0.3cm}
\subsection{Network Configurations}
All the three variants are configured according to the small version of offline SpatialNet (SpatialNet-small) \cite{quan_spatialnet_2024}, and the hidden size and the number of blocks are set to $H=96$ and $L=8$, respectively. For oSpatialNet-MSA, different window lengths \{2, 3, 4\} s are evaluated in our preliminary experiments, where the 4 s window performs slightly better and is used. For oSpatialNet-Ret, the RoPE is removed and four heads are used with exponential decay values set to $\gamma=$ \{$1-2^{-4}$, $1-2^{-5}$, $1-2^{-9}$, $1-2^{-10}$\} according to our preliminary experiments, where \{$1-2^{-4}$, $1-2^{-5}$\} and \{$1-2^{-9}$, $1-2^{-10}$\} are responsible for moving and static speakers, respectively. The negative of signal-to-noise ratio (SNR) is used as the loss function following \cite{ochiai_MaskBasedNeuralBeamforming_2023,wang_attention-driven_2023}.


STFT is applied using Hanning window with a length of 256 samples (32 ms) and a hop size of 128 samples (16 ms). For network training, the batch size is set to 4 utterances. For oSpatialNet-MSA and oSpatialNet-Ret, the Adam \cite{kingma2015adam} optimizer is used, while for oSpatialNet-Mamba the AdamW \cite{loshchilov_decoupled_2019} optimizer with a weight decay of 0.001 is used. The learning rate is initialized to 0.001 and exponentially decays as $lr \xleftarrow{} 0.001 * 0.99^{epoch}$. Gradient clipping is applied with a gradient norm threshold of 1.

\vspace{-0.3cm}
\subsection{Comparison of different training strategies}

\begin{figure}[tbp]
  \centering
  \includegraphics[width=1\linewidth]{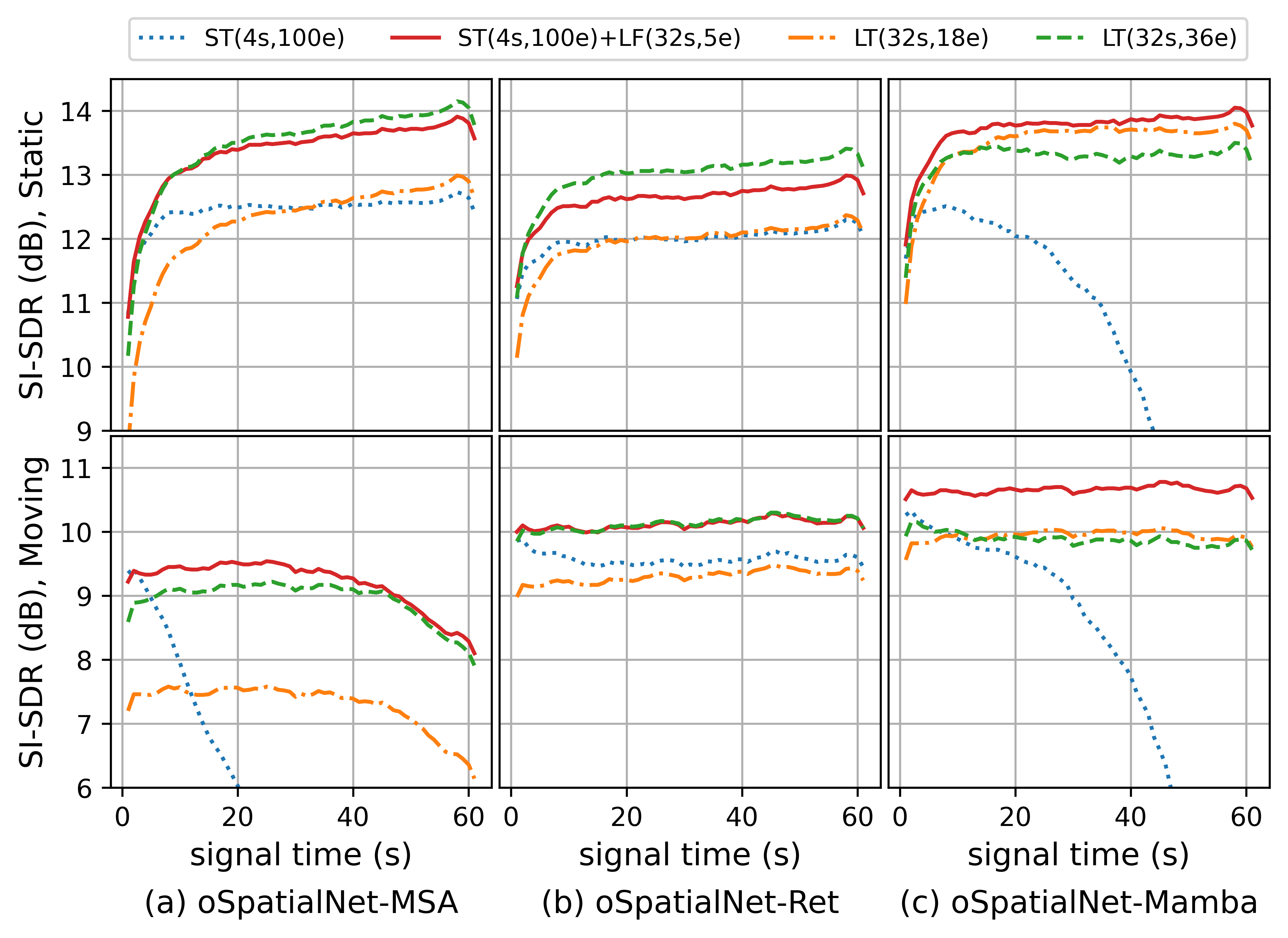}
   \vspace{-0.8cm}
  \caption{Speech enhancement performance for different training strategies. 
  }
  \label{fig_long_uttr}
  \vspace{-0.4cm}
\end{figure}

In this experiment, we compare three training strategies for long-signal inference,
i.e. short-signal training only (referred to as ST), short-signal training plus long-signal fine-tuning (referred to as ST+LF) and long-signal training only (referred to as LT).
Specifically, we trained the proposed three variants under four configurations denoted in the form of ``Strategy (training signal length, training epochs)'': 1) ST(4s,100e), 2) ST(4s,100e)+LF(32s,100e), 3) LT(32s,18e), and 4) LT(32s,36e).
Note that no matter how long the training utterances are, the same amount of training utterances and the same learning rate schedule are used.
As a result, ST(4s,100e)+LF(32s,5e) takes approximately the same training time with LT(32s,18e). \addnote[converge]{1}{LT(32s,36e) doubles the training time and the training process closely converges.}
The results on 64 s test utterances are presented in Fig. \ref{fig_long_uttr}, where the performance score of scale-invariant SDR (SI-SDR) \cite{roux_sdr_2019} is plotted and computed for every 4-s signal segments with a hop length of 1 s.

From Fig. \ref{fig_long_uttr}, we can see that:
1) Models trained with short signals usually have a good performance at the beginning stage. However, with the increase of signal time, longer historic context information provides limited performance improvement or even leads to performance collapse due to the length extrapolation problem in moving speaker cases for MSA and both static and moving speaker cases for Mamba.    
2) On top of short-signal training, after a few epochs of fine-tuning on long signals, the performance collapse of Mamba is fixed, and all models continuously improve their performance with the accumulation of longer-term context information for the static speaker case.
3) Taking the same training time as ST+LF, directly training with long signals for 18 epochs clearly shows some performance disadvantages than ST+LF. Doubling the training time to 36 epochs, the MSA and Retention models achieve similar performance with ST+LF, but the performance measures of Mamba get even worse. 

Overall, the proposed ST+LF training strategy first learns short-term speech enhancement knowledge and then extend the knowledge to long-term processing, which is efficient in terms of both training cost and speech enhancement performance. 

\begin{table}[tpb]
\setlength\tabcolsep{2pt}
\renewcommand{\arraystretch}{1.0}
\caption{Multichannel speech denoising and dereverberation results.}
\vspace{-0.2cm}
\label{table_results}
\centering
\resizebox{\linewidth}{!}{
\begin{tabular}{lccccccc}
\hline\hline
Network & \#Param & FLOPs & Causal & NB-PESQ & ESTOI & SI-SDR   & SDR  \\
Architecture &  (M) & (G/s) &  &      &       &  (dB)    & (dB)  \\
\hline\hline
\multicolumn{8}{c}{\textbf{Results for static-speaker cases
}}\\
\hline
unproc. & - & - & - & 1.62 & 0.350 & -8.2 & 0.5 \\ %
\hline
EaBNet & 2.4 & 4.4 & \cmark & 2.68 & 0.744 & 5.4 & 8.5 \\
EaBNet$^+$ & 9.6 & 17.4 & \cmark & 2.79 & 0.770 & 6.5 & 9.3 \\
McNet & 1.9 & 29.7 & \cmark & 2.82 & 0.769 & 7.2 & 9.6 \\ %
\textcolor{gray}{SpatialNet} & \textcolor{gray}{1.2} & \textcolor{gray}{23.1} & \textcolor{gray}{\xmark} & \textcolor{gray}{3.55} & \textcolor{gray}{0.921} & \textcolor{gray}{15.2} & \textcolor{gray}{17.2} \\ %
\hline
oSpatialNet-MSA & 1.2 & 23.1 & \cmark & 3.21 & 0.869 & 13.4 & 14.8 \\ %
oSpatialNet-Ret & 1.4 & 20.4 & \cmark & 3.11 & 0.852 & 12.6 & 14.0 \\ %
oSpatialNet-Mamba & 1.4 & 18.4 & \cmark & \textbf{3.27} & \textbf{0.880} & \textbf{13.7} & \textbf{15.2}  \\ %
\hline\hline
\multicolumn{8}{c}{\textbf{Results for Moving-speaker cases
}}\\
\hline
unproc. & - & - & - & 1.60 & 0.340 & -8.4 & -2.3 \\ %
\hline
EaBNet & 2.4 & 4.4 & \cmark & 2.64 & 0.732 & 4.7 & 7.2 \\
EaBNet$^+$ & 9.6 & 17.4 & \cmark & 2.75 & 0.757 & 5.7 & 8.0 \\
McNet & 1.9 & 29.7 & \cmark & 2.70 & 0.737 & 5.4 & 7.7 \\ %
\textcolor{gray}{SpatialNet} & \textcolor{gray}{1.2} & \textcolor{gray}{23.1} & \textcolor{gray}{\xmark} & \textcolor{gray}{3.20} & \textcolor{gray}{0.869} & \textcolor{gray}{11.4} & \textcolor{gray}{13.0} \\ %
\hline
oSpatialNet-MSA & 1.2 & 23.1 & \cmark & 2.88 & 0.797 & 9.2 & 10.8 \\ %
oSpatialNet-Ret & 1.4 & 20.4 & \cmark & 2.94 & 0.816 & 10.1 & 11.6 \\ %
oSpatialNet-Mamba & 1.4 & 18.4 & \cmark & \textbf{3.05} & \textbf{0.836} & \textbf{10.7} & \textbf{12.2} \\ %
\hline\hline
\end{tabular}
}
\vspace{-0.4cm}
\end{table}

\vspace{-0.4cm}
\subsection{Comparison of the three models}

The proposed oSpatialNet mainly exploits spatial information, such as the difference of spatial correlation between directional speech and diffuse ambient noise for denoising, and the (narrow-band) RIR information for dereverberation. Some spatial information are time-invariant (for the entire test signal) such as the spatial correlation of signals and the RIR for static speakers, while some others are time-variant such as the RIR for moving speakers. Different models learn these long-term and short-term information in different ways. MSA and Retention are both self-attention models, and use a rectangular window and an exponentially decaying window to cut/forget the past sequence, respectively. Even though the window length could be small, they are still able to exploit longer-term context via stacked layers. MSA treats the time steps within the memory indiscriminately, while Retention gives larger weights to more recent time steps. As a result, MSA performs better for the static speaker case by exploiting longer-term context information, while Retention performs better for the moving speaker case by relying on more recent context information. In addition, by exponentially forgetting historic information, Retention has a good length extrapolation performance even when trained with only short signals. Mamba is an RNN-like model, it memorizes and compresses useful historic information in its state space when using input-dependent selective parameters. Mamba (with ST+LF) works better for both the static and moving speaker cases than MSA and Retention, which indicates that the selective parameters are indeed able to adaptively select useful data.



\vspace{-0.4cm}
\subsection{Comparison with baseline methods}

In table \ref{table_results}, the results of the proposed oSpatialNet are compared with three baseline methods, i.e. an advanced all-neural beamformer EaBNet \cite{li_EmbeddingBeamformingAllNeural_2022}, an advanced multi-cue fusion network McNet \cite{yang2022mcnet} and the offline SpatialNet \cite{quan_spatialnet_2024} (small version). 
Note that, for EaBNet, a version denoted as EaBNet$^+$ with twice the hidden units reported in its paper is also compared.
For fair comparison, McNet, EaBNet and EaBNet$^+$ are also trained with ST+LF, which improves their SI-SDR performance compared to ST only about 0.3, 2.5 and 1.4 dB respectively.
The offline SpatialNet is trained with 4-s utterances for 100 epochs. During test, it processes long utterances by chunking them to 4-s utterances with 2-s of overlapping.
For each method, the number of model parameters (\#Param), the number of floating point operations (FLOPs), and speech enhancement performance for both static and moving speaker cases are reported.  Four performance metrics are used, including narrow-band perceptual evaluation of speech quality (NB-PESQ) \cite{rix_perceptual_2001}, extended STOI (ESTOI) \cite{jensen2016algorithm}, SDR \cite{vincent_performance_2006} and SI-SDR. Performance are tested on 64-s signals, and the performance scores are computed for every 4-s segments with 1-s hop length, and then averaged over all segments. From the table, we can see that the proposed oSpatialNets all largely outperform the online McNet, EaBNet and EaBNet$^+$. 
Among the three variants, oSpatialNet-Mamba performs the best.
\addnote[noncausal]{1}{As expected, the offline SpatialNet outperforms all the online methods by leveraging future information.}

\vspace{-0.3cm}
\section{Conclusion}
This paper proposes the online SpatialNet for long-term streaming speech denoising and dereverberation. Three variants have been developed using the networks of masked self-attention, Retention and Mamba, respectively. In addition, a short-signal training and long-signal fine-tuning strategy is proposed to improve the length extrapolation ability of the networks within limited training time. The proposed online SpatialNet, especially the Mamba variant, achieves outstanding performance for long-term streaming speech enhancement and for both static and moving speakers.








\bibliographystyle{IEEEtran}
\footnotesize
\bibliography{mybib}

\end{document}